\documentclass[10pt,spanish]{article}

\usepackage[spanish,activeacute]{babel}
\usepackage[latin1,utf8]{inputenc}
\usepackage{latexsym}
\usepackage{indentfirst}

\usepackage{natbib}
\bibpunct{(}{)}{;}{a}{}{,} 

\usepackage{txfonts}
\usepackage{times}

\usepackage{amssymb}
\usepackage[hyphens]{url}
\usepackage{lscape}

\usepackage[usenames,dvipsnames]{color}
\definecolor{lightblue}{rgb}{0,0.3,0.9}
\usepackage{lastpage}

\begin{document}

\centerline{\bf{\LARGE{Observational and Theoretical study of the}}}  \centerline{\bf{\LARGE{inner region of HH~30}}}

\vspace{0.3in}
\centerline{\bf Phd thesis}
\centerline{Author: Maria Carolina Duran Rojas}

\centerline{Thesis Advisor: Alan Watson}

\vspace{0.5in}
\centerline{\large \bf Abstract}

The theory of low mass young stellar formation establish that they 
form because of the gravitational contraction of the molecular cloud. 
The cocoons, where stars will form, rotate. This rotation causes the 
formation of a keplerian disk around the star. The disk is composed 
of gas and dust. Observationally, in the optical and near infrared 
wavelengths, is very easy to see sources with edge-on disks, because 
they are optically thick and veil the light from the star. The images of 
edge-on disks consist of the scattered starlight by the dust in the 
disk in the infrared. On the other hand, in the optical wavelengths, 
is more complicated to observe sources with a disk that is parallel
to the plane of the sky because we are looking directly at the star.

HH 30 is a classical T Tauri star that has been extensively studied 
during the last two decades. Hubble Space Telescope ({\it HST}) 
observations reveal the existence of a variability in the external 
disk illumination, this is the result of the starlight scattered by 
the dust in the disk. In this study our goal is to determine the 
period or periods of the variability or better constrain the time 
scale of a period. For example, photometric studies of the outer 
disk variability in HH 30 indicate that this periods must be less  
than 6 months, some other  suggest  that  they  must  be  less
than 300 days. It is important to note that depending on the 
period scale, the variability should related with the star or with 
the disk. The mechanisms related with short periods are  
connectedwith the photosphere or regions close to the star,  
however, mechanisms linked with long periods are related to 
asymmetries in the inner parts of the accretion disk. Previous 
photometric studies of the outer disk variability in HH 30 indicate 
that this periods must be less than 6 months. It is important to 
note that the period may indicate whether the variability should 
related with the star or with the disk. The mechanisms related 
with short periods are connected with the photosphere or regions 
close to the star. However, mechanisms linked with long periods 
are likely to be asymmetries in the inner parts of the accretion disk.
To fulfill our goal, we made an observational and theoretical 
analysis of HH 30. At first we obtained direct photopolarimetric
images with a the 84 cm of National Observatory of Mexico located
at San Pedro Martir (OAN-SPM). We obtained images  of  the  
object  in  four different  positions  of  the  analyzing  prism.  
This different positions allow us to determine the normalized 
Stokes parameters {\it q} and {\it u}, which reveal variability. 
We tried to measure the photomeric variability, howeverthis 
was not periodic.

A method used to determine periods is called the Lomb-Scargle 
normalized periodogram, which calculates a power spectral 
density, that depends on the observable quantities like the 
magnitude and the observation time. We applied this method 
to three sets of photometric data, but we do not find a 
variability period of the source. It drives us to look for a different 
way to mitigate short time correlations effects, which consist of 
binning the sets of data in a 1/8 of the period that we want to 
test. We calculated the Lomb-Scargle periodogram of the 
polarimetric  data and we found a variability period of 7.5 days. 
The level of significance of this polarimetric period was high 
whereas the significance of the same period in the photometry 
was lower. However, we found that both periods are consistent,
and have a particular characteristic, that the sin fit of photometric 
variability is displaced by a quarter of the sin fit of polarimetric 
variability. That means that the photometric component shows 
a minimum or a maximum when the polarimetric component is 
null.

Different mechanisms have been suggested to explain the 
variable asymmetry, including  mechanisms  that are directly  
linked to the star and mechanisms that are not.  In the first 
place, in the photosphere there are hot  spots that are 
produced  by  the  shock of material from the disk. In the 
other hand, they are mechanisms that are related with
asymmetries in the inner disk or a close companion like a 
brown dwarf or a big planet. Our result that sin fits of 
photometric and polarimetric variability are displaced by 
a quarter in between are in agreement to the mechanisms 
related directly to the star, like the hot spots. A model 
proposed to explain the variability is the lighthouse model. 
This model consists of a beam or a shadow from the central 
source that has an azimuthal movement that sweep the full 
disk. The mechanism proposed for the lighthouse model are, 
the none symmetric accretion produced by hot spots, and 
clumps or gaps in the disk. Unfortunately, however we 
determine a variability period we were not able to distinguish 
between the mechanisms but our observations provide
quantitative constrains of the period and in the photometric 
and polarimetric modulation amplitude.

We show that one of the mechanisms proposed to explain the 
variability for HH 30 can reproduce our polarimetric observations. 
One model include a low mass (2 solar masses) star with hot 
spots in its photosphere and a flared optically thick disk round 
the source. We obtained the intensity of HH 30 with the radiative 
transfer code of Watson y Henney (2001) that includes polarization. 
We used the parameters of Wood y Whitney (1998) to calculate the 
polarimetric variability. The important parameters of that model 
for the source are mass, radius and temperature, mass, size, 
inclination, flaring, the exponent of the density power-law,
size, altitude and brightness of the hot spots, the phase function 
for scattering of Henyey-Greenstein, albedo and opacity for the 
dust. We discussed these parameters and suggest a model that can 
reproduce our polarimetric observations.

\end{document}